# The Role of Phase Stabilization and Surface Orientation in 4,4'-Biphenyl-Dicarboxylic Acid Self-Assembly and Transformation on Silver Substrates

*Anton O. Makoveev,[1] Pavel Procházka,[1] Matthias Blatnik,[1] Lukáš Kormoš,[1] Tomáš Skála,[2] and Jan Čechal[1,3]\**

[1] CEITEC – Central European Institute of Technology, Brno University of Technology, Purkyňova 123, 612 00 Brno, Czech Republic.

[2] Department of Surface and Plasma Science, Faculty of Mathematics and Physics, Charles University, V Holešovičkách 2, 18000 Prague 8, Czech Republic

[3] Institute of Physical Engineering, Brno University of Technology, Technická 2896/2, 616 69 Brno, Czech Republic.

AUTHOR INFORMATION

**Corresponding Author**

\* E-mail: cechal@fme.vutbr.cz (J. Č.)

ABSTRACT

Molecular functionalization of nanoparticles and metallic substrates can be used to tune their properties for specific applications. However, polycrystalline substrates and nanoparticles exhibit surface planes with distinct crystallographic orientations. Therefore, the development of reliable strategies for molecular functionalization requires knowledge of the role of the surface plane orientation in the growth kinetics, structure, and properties of the molecular layer. Here, we apply a combination of low-energy electron microscopy and diffraction, scanning tunneling microscopy and photoelectron spectroscopy to investigate the self-assembly of 4,4'-biphenyl-dicarboxylic acid (BDA) on Ag(111) and critically discuss the difference to Ag(100). Whereas the structural motifs for intact and fully deprotonated BDA are similar on both surfaces, the intermediate phases comprising partially deprotonated BDA differ in the structure and chemical composition. A real-time view of the phase transformations enables us to present a generalized picture of the phase transformations between the self-assembled molecular phases on the surfaces and underline important features such as the phase stabilization of the chemical composition and the mechanism of the related burst transformation. The influence of the substrate orientation on the structure of molecular layers and phase transformations provides the necessary background for developing functionalization strategies of the substrates displaying multiple surface planes and kinetic models for the growth near thermodynamic equilibrium.

INTRODUCTION

Molecular self-assembly on surfaces is a bottom-up approach for constructing atomically precise two-dimensional structures, providing unusual and even unique properties.[1–6] Diverse on-surface self-assembled architectures have already been created and tested, e.g., as catalysts and host-guest systems.[7–14] However, applications of these architectures are still hindered, mainly due to large differences in structure and properties of molecular layers on substrates from different materials, with different crystallographic orientation, or even with different step densities, because self-assembly is governed by a subtle interplay of involved intermolecular and molecule-substrate interactions.[4,15,16] In this regard, "model" molecular systems on single-crystal substrates serve as a vital source of information, i.e., for the functionalization of nanoparticle surfaces exhibiting multiple facets[17,18] or polycrystalline metal sheets (e.g., electrodes), which requires the knowledge of molecular self-assembly on surfaces with various surface orientations.

Both intermolecular and molecule-substrate interactions play a significant role on the molecular layer- solid surface interface. Depending on the structure and chemical composition of adsorbed molecules and the structure and electronic properties of the surface, distinct scenarios of the molecule-substrate interface formation are realized.[19] In the case of planar aromatic molecules and an inert surface, a flat-lying adsorption geometry occurs to maximize the contact area between the molecule and substrate, coupled predominantly by van der Waals interactions. The presence of functional groups dramatically changes both intermolecular- and molecule-substrate interactions, which results in distinct bonding patterns, induction of the structural (reconstruction)[20,21] and morphological (reshaping),[22,23] changes to the substrate, or incorporation of substrate adatoms into bonding patterns[24]. At the molecule-substrate interface, interactions with no analogs in the bulk or liquid may exist and have a dominant effect on the resulting structure.[21,25,26] An interplay between molecule-molecule and molecule-substrate interactions turns out to be decisive for the resulting arrangement of the employed molecules.[4] The term 'subtle interplay' expresses that even small changes in the substrate or molecule can cause substantial alterations in the resulting self-assembled structure.[16]

A good example of altered self-assembly can be found at different facets of the same material. The separate studies on self-assembly of 4,4′-biphenyl dicarboxylic acid (BDA) on Cu(111) and Cu(001) show very different behavior of BDA on these substrates: the molecular phases have different structures, the deprotonation of carboxyl groups occurs at different temperatures, and nucleation of the phases follows a different mechanism.[27,28] The self-assembly of tetracyano-p-quinodimethane (TCNQ) on Cu(111) and Cu(100) also demonstrates remarkable dissimilarities both in structure and morphology. The deposition of TCNQ on Cu(111) at room temperature leads to the formation of small ordered domains together with disordered regions,[29] whereas compact islands with well-defined rectangular shapes grow on Cu(100) at the same deposition conditions.[21] Another investigation has reported significant differences in the self-assembly of octaethylporphyrin on surfaces of different materials (Cu(100), Ag(100), and Au(100)) and also of different orientations (Ag(100) and Ag(110)), which considerably affect the on-surface molecular dynamics and reactivity,[30] signifying the importance of the substrate structure. Although there are major steps towards understanding the facet role in molecular self-assembly at surfaces for various systems, this knowledge is still rather scarce.

In this respect, we present a study on the self-assembly of BDA (Figure 1) on Ag(111) with a detailed comparison of its behavior on Ag(100). BDA presents a well-known model molecule for the synthesis of multiple self-assembled architectures at diverse substrates.[27,28,31–38] In this work, we utilize a set of complementary experimental methods, i.e., scanning tunneling microscopy (STM), low-energy electron microscopy/diffraction (LEEM/LEED), and X-ray photoelectron spectroscopy (XPS), to describe the BDA phases self-assembled on a Ag(111) surface. The thermally induced chemical change of BDA, i.e., the deprotonation of its carboxyl groups, results in the formation of several molecular phases, each showing a different structure and ratio of intact, singly- and fully- deprotonated BDA molecules. The detailed comparison with BDA on Ag(100)[31,32,38] allows description of the role of the substrate structure on the BDA self-assembly and phase transformations between the individual phases.

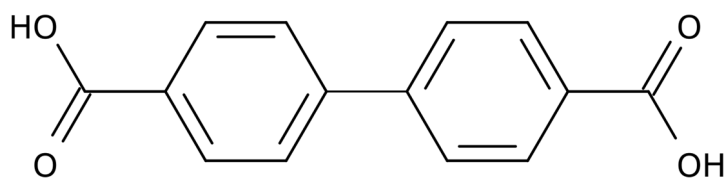

**Figure 1.** The structural formula of a BDA molecule.

METHODS

Experimental data presented in this study were collected in a multi-chamber ultrahigh vacuum (UHV) cluster at the CEITEC Nano Research Infrastructure in Brno, Czech Republic, and at the Material Science beamline of the Elettra synchrotron light source in Trieste, Italy. The UHV cluster consists of 9 separable UHV chambers connected via a transfer line maintaining UHV conditions (base pressure $2\times10^{-10}$ mbar), and 5 of these have been used in this study to prepare and characterize samples. During the movement of samples in-situ from one chamber to the next, an increase in pressure of up to $2\times10^{-9}$ mbar is observed, but upon ceasing of the movement, the pressure recovers within seconds. Details of the UHV chambers are listed below, following the sample handling during measurement.

The sample, an Ag single crystal (MaTeck or SPL), cut along the (111) plane, was prepared in one of the UHV chambers dedicated to sample cleaning ("Preparation chamber") with a base pressure of $2\times10^{-10}$ mbar. The crystal was cleaned by $Ar^+$ sputtering and subsequent annealing to 803 K in UHV followed by slow cooling (< 0.5 K/s) to 300 K. After cleaning, the sample was transferred in-situ via the UHV transfer line to another UHV chamber dedicated to the deposition of atoms and molecules ("Deposition Chamber") with a base pressure of $8\times10^{-11}$ mbar. Therein, BDA molecules, purchased from Sigma–Aldrich (purity 97 %), were deposited by molecular beam epitaxy (MBE) on the sample, which was held at room temperature, by a near-ambient effusion cell (Createc) from an oil-heated crucible held at 438 K before usage for several hours to ensure their purity. A surface fully covered by molecules was obtained by BDA deposition for 20 min at a pressure lower than $8\times10^{-10}$ mbar. We define 1 ML of BDA coverage as the crystal surface fully covered with the so-called α phase, i.e., ~0.1 molecules per substrate unit cell area.

Following deposition, LEEM and LEED measurements were carried out in a UHV-connected customary Specs FE-LEEM P90 instrument where a base pressure of $2\times10^{-10}$ mbar can be achieved. Diffraction patterns were collected during measurement from a surface area of $15\times10$ $\mu m^2$ and bright-field images were formed by detecting electrons with energies of up to 7 eV from the (0,0) diffracted beam. Microdiffraction patterns were obtained from an area with a diameter of 185 nm. In the case of elongated diffraction spots, we have taken the most intense regions of those spots to define the reciprocal unit cells. These most intense regions approximately correspond to the center of the respective diffraction spots.

STM images were collected in another, separate, UHV chamber with a commercial Specs Aarhus 150 system at a base pressure of $1\times10^{-10}$ mbar. The STM is equipped with electrochemically etched tungsten tips. The images shown in this work were all measured at 300 K (room temperature) in constant current mode. The tunneling current varied between 60 and 100 pA. The bias voltage was set to negative bias values between -150 and -300 meV; imaging filled states. Distortions due to lateral thermal drift at 300 K were corrected by a linear drift obtained from a series of consecutive images.

Finally, the fifth UHV chamber holds a commercially available system able to collect XPS as well as UV Photoemission Spectroscopy (UPS, not shown in this study). Data were collected with a Phoibos 150 spectrometer in normal emission geometry (NE: emission angle 0°) and a 2D CCD detector (both Specs) by shining X-ray light from a non-monochromatized Mg K$\alpha$ source. Detailed core level (CL) spectra were measured in medium magnification mode, integrating 30 sweeps with 0.1 s dwell time, 0.05 eV energy step width, and 20 eV pass energy. The total energy resolution (combining analyzer and excitation radiation contributions) was 800 meV. The obtained spectra were fitted employing Voigt-shaped profiles and a Shirley (Shirley + parabolic) background. The availability of the *in-situ* XPS device assures that the data have been measured from the same sample as measured by LEEM and STM featuring the particular BDA phase.

Synchrotron radiation photoemission spectroscopy (SRPES) performed at the Materials Science Beamline at the Elettra synchrotron light source in Trieste is used to collect data with a better signal-to-noise ratio, especially regarding the detailed core-level spectra mentioned above. Excitation energies of 420, 510, and 620 eV for C 1s, Ag 3d, and O 1s core levels, respectively, have been used in this regard. The spectra were acquired in medium area lens mode using 10 eV pass energy. The resulting spectrum was obtained by integrating 3 (Ag 3d, C 1s) or 25 (O 1s) sweeps taken with 0.1 s dwell time and 0.05 eV energy step. The total resolution achieved was in the range of 300–550 meV. Peak positions were corrected with respect to the Fermi edge of the Ag substrate measured for each excitation energy, and the spectra were normalized with respect to the intensity of the photo-current measured on a gold mesh placed in the beamline. The temperature was measured by a K-type thermocouple attached to the bottom side of the Ta sample plate (thickness: 0.1 mm). The collected datasets were fitted with single/multiple components using Voigt-shaped profiles after a Shirley background subtraction, as demonstrated in Supplementary Information, Section 2.

RESULTS AND DISCUSSION

In this study, we thoroughly investigate BDA molecular phases and their transformations on a Ag(111) substrate and compare them to those on Ag(100). This enables us to describe the influence of the substrate's crystallographic orientation on BDA self-assembly. To prevent confusion, we explicitly give the substrate orientation to the index of phase denomination. Upon annealing, BDA on Ag(111) subsequently forms four different phases, namely $\alpha^{(111)}$, $\alpha_s^{(111)}$, $\beta^{(111)}$, and $\delta^{(111)}$, characterized by their structure and ratios of fully protonated, semi-, and fully deprotonated BDA molecules evidenced by XPS (Figure 2). A detailed overview of their morphology, diffraction, and molecular models can be found in Figure 3, and their characteristics are summarized in Table 1. On Ag(100), the gradual deprotonation of BDA also leads to the formation of distinct molecular phases: $\alpha^{(100)}$, $\dot{\alpha}^{(100)}$, $\beta^{(100)}$, $\gamma^{(100)}$, and $\delta^{(100)}$, which have been thoroughly described in our previous works.[31,32,38] We show clear distinctions between analogous molecular phases of the two considered substrates and, in the following subsections, provide a detailed description of BDA phases on Ag(111) and their comparison with Ag(100).

**Table 1**. Characteristics of the BDA phases on Ag(111).

| BDA Molecular Phase | Degree of deprotonation (%)[a] | Mean Island Size [µm²] and Island Shape[b] | Number of Symmetry Equivalent Domains[c] | Unit Cell[d,e] |
|---|---|---|---|---|
| $\alpha^{(111)}$ | 0 | 0.51×0.06 elongated | 6 | – |
| $\alpha_s^{(111)}$ | 10 – 30 | 0.76×0.17 elongated | 6 | $\begin{pmatrix} \frac{34}{9} & \frac{95}{63} \\ 0 & \frac{18}{7} \end{pmatrix}$ |
| $\beta^{(111)}$ | ~63 | 1.1 round | 6 | $\begin{pmatrix} 8 & \frac{96}{11} \\ -3 & \frac{28}{11} \end{pmatrix}$ |
| $\delta^{(111)}$ | 100 | >2.1 round | 3 | $\begin{pmatrix} 4 & 2 \\ 0 & 6 \end{pmatrix}$ |

Notes:

[a] The degree of deprotonation was determined from the XPS measurements detailed in Supplementary Information, Sections 1 and 2.

[b] The mean island size was defined from the bright-field LEEM images, examples of which are given in Figures 3a-d.

[c] The number of symmetry equivalent domains and molecular unit cells were determined from the diffraction patterns given in Figures 3e-h detailed in Supplementary Information, Sections 3, 4, 5, and 6.

[d] The molecular unit cells were identified by the local congruence method[31,32] and are expressed in matrix notation.

[e] We have found two distinct moiré structures for the $\beta^{(111)}$ phase (see Supplementary Information, Section 5 for details).

# The α$^{(111)}$ and α$_s^{(111)}$ phases

A 0.6 – 0.8 monolayer (ML) coverage of the α$^{(111)}$ was prepared by the deposition of BDA on the substrate held at room temperature. The α$^{(111)}$ phase contains only fully protonated BDA molecules, as confirmed by the XPS analysis (Figure 2, see details in Supplementary Information, Section 1). The STM image in Figure 3i shows BDA as a bright rodlike feature with an average length (at 50 % of apparent height), width, and apparent height equal to 0.96 ± 0.05 nm, 0.42 ± 0.02 nm, and ~50 pm, respectively. The molecules are organized in straight molecular rows of complementary hydrogen-bonded molecules (indicated by the black arrow in Figure 3i) with a mutual distance of 0.59 ± 0.01 nm. The molecular rows are tilted from the principal surface directions by ~12°, as demonstrated in a tentative model in Figure 3m. Bright-field LEEM imaging uncovers bright needle-like BDA islands on a dark Ag(111) substrate. The islands seem to grow out of the step edges and are oriented in the same direction as the molecular rows comprising them (Figure 3a and Supplementary Information, Section 3). This is remarkably similar to the α$^{(100)}$, where BDA molecular islands represent elongated entities with preferential growth starting at the substrate's step edges. The microdiffraction shown in Figure 3e demonstrates that the structure of the α$^{(111)}$ phase is periodic, which is in striking contrast to BDA on Ag(100), where the α$^{(100)}$ is not periodic within the molecular rows.[31] Compared to Ag(100), where the inter-row periodicity exactly matches 2 substrate atom distances, the Ag(111) surface is more densely packed. Therefore, the molecular rows no longer precisely fit the substrate, which results in a slight variation of the inter-row periodicity expressed as the elongation of the diffraction spots.

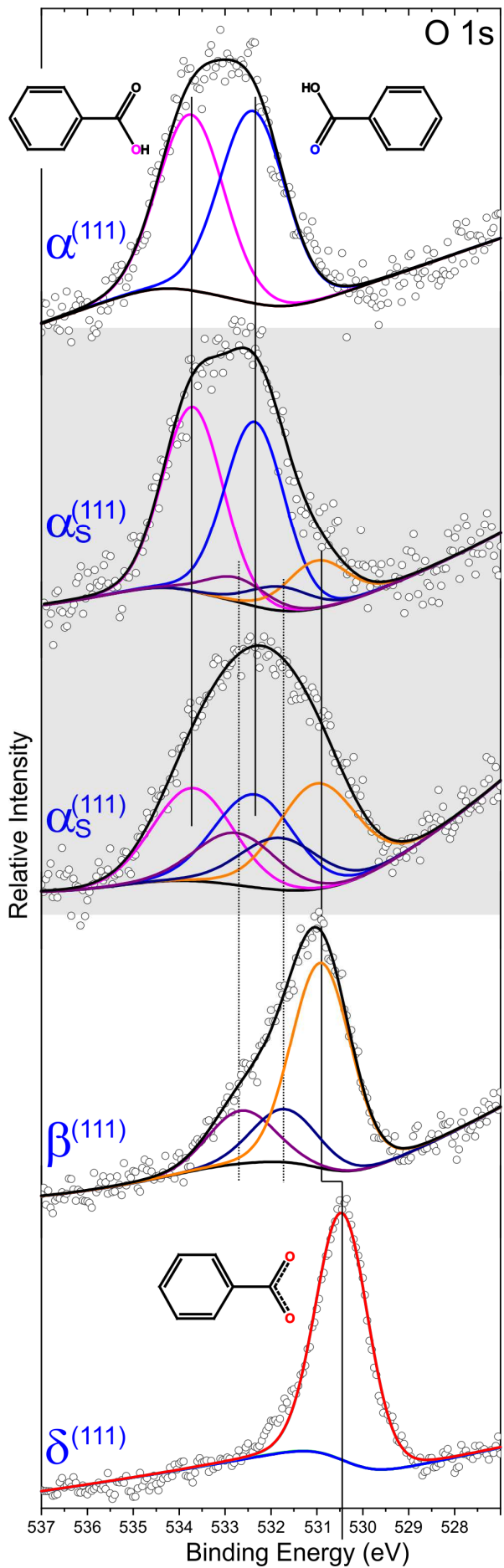

**Figure 2.** XPS of BDA phases on Ag(111) measured on samples with a given phase confirmed by LEEM. A detailed description of the single peak components used to fit the obtained spectra can be found in the text and Supplementary Information, Section 1. The two spectra ascribed to the $\alpha_s^{(111)}$ reflect its variable degree of deprotonation; the annealing in LEEM was stopped either immediately after the formation of $\alpha_s^{(111)}$ phase (324 K, 5 min) or after 45 min at 340 K before it transformed into the $\beta^{(111)}$.

The annealing of the Ag(111) surface with the $\alpha^{(111)}$ phase at 373 K results in changes in island morphology, diffraction pattern, and chemical composition. Whereas the position of diffraction spots and overall look of the microdiffraction pattern (Figure 3f) of the new phase resembles the $\alpha^{(111)}$, the diffraction spots are circular and sharper. As the sharp diffraction spots are the dominant feature of this phase, we name it $\alpha_s^{(111)}$. In comparison to the $\alpha^{(111)}$, molecular islands of the $\alpha_s^{(111)}$ are larger (see Table 1), as shown in the bright-field LEEM image given in Figure 3b. The additional diffraction spots (highlighted by the black arrows in Figure 3f) are due to the moiré pattern associated with the full $\alpha_s^{(111)}$ unit cell commensurate with the substrate (see Supplementary Information, Section 4 for details). The STM data in Figure 3j also demonstrate that the real-space structure of the $\alpha_s^{(111)}$ is remarkably similar to the $\alpha^{(111)}$ (for large-scale STM of the $\alpha^{(111)}$ and $\alpha_s^{(111)}$ see Figure S4.2).

A notable difference between the $\alpha^{(111)}$ and $\alpha_s^{(111)}$ is revealed by XPS. The XPS spectra shown in Figure 2 depict two possible compositions of the $\alpha_s^{(111)}$ with different degrees of BDA deprotonation. In addition to the peaks associated with the $\alpha^{(111)}$ phase, the peaks associated with the carboxyl-carboxylate motif are present (Supplementary Information, Section 1), in accordance with the analogous phases $\dot{\alpha}^{(100)}$ and $\beta^{(100)}$ on Ag(100). From the very similar structure and variable chemical composition, we deduce that the deprotonated molecules can be accommodated in the molecular islands of the $\alpha^{(111)}$, first as defects, and at a higher concentration they induce the formation of the $\alpha_s^{(111)}$ in which they are a structural unit. This is the same behavior we have observed for the $\dot{\alpha}^{(100)}$. The deprotonated carboxyl groups provide a stronger binding to the substrate, which makes the influence of the substrate significant. By analogy to Ag(100), enforcing a strict substrate periodicity explains the sharper diffraction spots.[38]

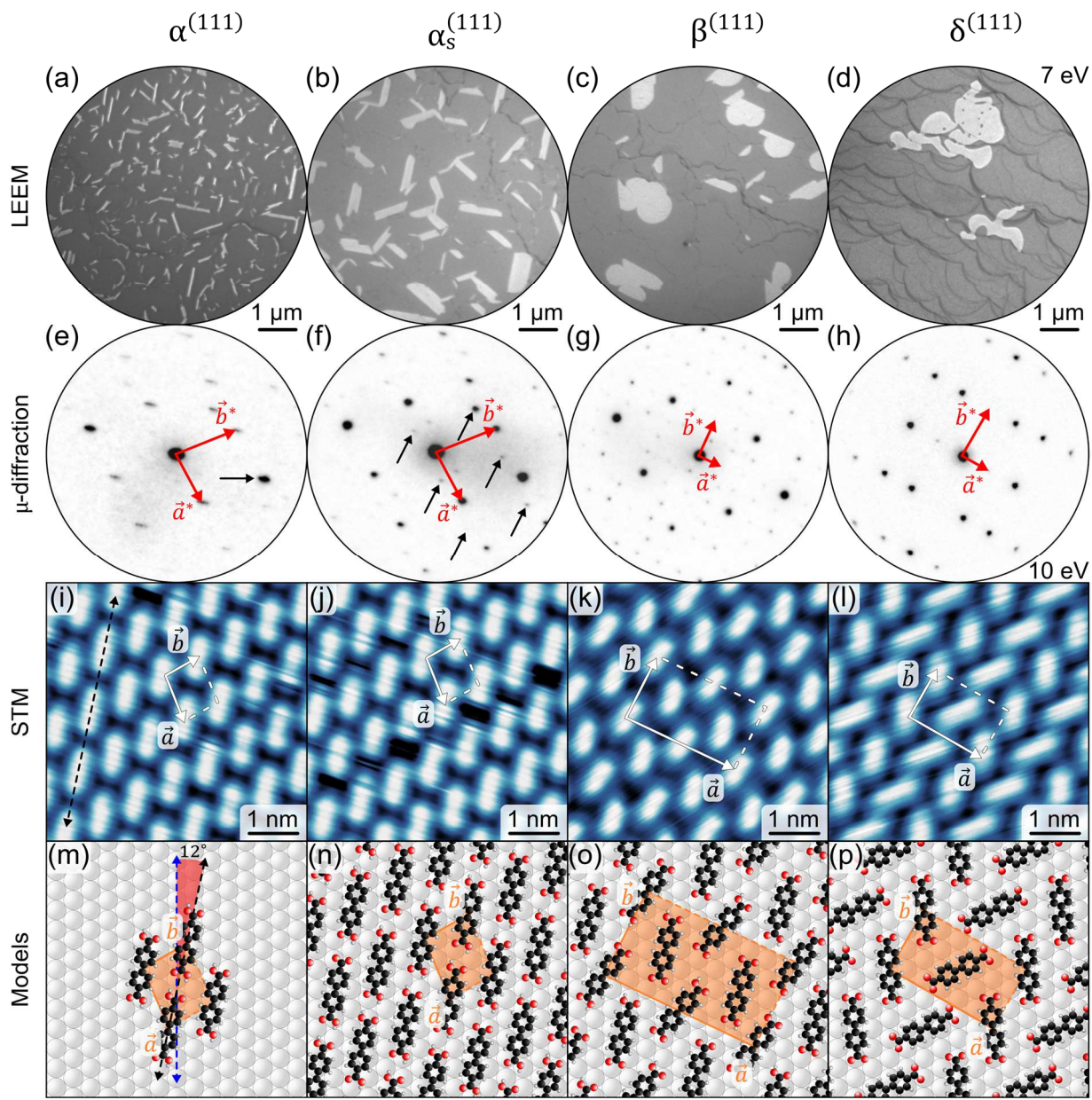

**Figure 3.** BDA phases on Ag(111). (a-d) Bright-field LEEM images of all investigated phases measured at 7 eV; (e-h) Microdiffraction patterns measured at 10 eV primary energy. The black arrow in (e) shows the spot associated with the inter-row periodicity of the $\alpha^{(111)}$ (see text for details). The red arrows indicate the reciprocal unit cell vectors. The black arrows in (f) highlight the diffraction spots associated with the $\alpha_s^{(111)}$ moiré pattern. (i-l) STM images of the different BDA phases; real-space unit cell vectors are depicted by the white arrows. The black dashed double-sided arrow in (i) demonstrates the direction of the $\alpha^{(111)}$ molecular rows. (m-p) Model representations of the phases. The unit cells are highlighted by the orange rhomboids.

The blue and black dashed double-sided arrows in (m) indicate the angle between the $\alpha^{(111)}$ molecular rows and the principal surface directions. All the measured data were aligned with respect to the corresponding models in (m-p).

## The $\beta^{(111)}$ and $\delta^{(111)}$ phases

The $\beta^{(111)}$ phase appears when the sample is annealed at ~400 K. Bright-field LEEM measurement shows the round-shaped islands of the $\beta^{(111)}$, each extending over several terraces (Figure 3c). Compared to the $\alpha_s^{(111)}$, the islands are located further away from each other. The STM image in Figure 3k points to an interlocked row molecular configuration with alternating planar and buckled BDA (see the model in Figure 3o), similarly to the $\beta^{(100)}$ on Ag(100).[31] The chemical composition investigated by XPS includes ~63 % of oxygen incorporated in carboxylates (Figure 2 and Supplementary Information, Section 1), which approximately corresponds to 5 deprotonated carboxyl groups out of 8 in the unit cell. The shifts observed for the peaks associated with the oxygen atoms closely resemble those for the $\beta^{(100)}$.[31] Whereas the chemical state of the $\beta^{(111)}$ can be compared to the $\gamma^{(100)}$ phase comprising ~66 % carboxylates, its structure is close to $\beta^{(100)}$.

The fully deprotonated, $\delta^{(111)}$, phase on Ag(111) can be prepared by annealing the sample at 453 K. Our XPS analysis of the $\delta^{(111)}$ identifies a single peak at 530.45 eV that can be associated with carboxylate oxygen (red in Figure 2). The XPS measurements also detect a decrease in the signal of both carbon and oxygen caused by the desorption of part of the molecules from the substrate. Bright-field LEEM imaging (Figure 3d) reveals large molecular islands with an average size >2.1 µm², the biggest single domain BDA islands we have ever observed on both Ag(111) and Ag(100) surfaces at coverages below 1 ML. The structure of the

$\delta^{(111)}$ phase is defined by the binding motif where deprotonated carboxyl groups point towards the phenyl ring of neighboring BDA molecules (Figure 3l) in a unit cell commensurate with the substrate (Supplementary Information, Section 6). The structure is essentially the same as on Ag(100).[32]

**The role of substrate in self-assembly**

The BDA deprotonation is significantly influenced by substrate material and surface plane orientation. It has been shown that BDA on Cu(100) forms a self-assembled structure consisting of fully deprotonated molecules already at room temperature.[28] On Cu(111), the situation is slightly different:[27] a mixture of partially deprotonated and fully deprotonated phases appears on the surface at room temperature. This indicates a lower reactivity of the (111) surface compared to the (100). However, in general, Cu has a considerably higher reactivity for deprotonation of BDA than Ag and Au. Indeed, in our previous studies,[31,32,38] we have shown that BDA on Ag(100) deprotonates very slowly at room temperature, and annealing at higher temperatures is necessary to reach a noticeable degree of the deprotonation in a reasonable time, which is in agreement with works on carboxylic acids by other groups.[39–41] A Ag(111) surface is less reactive with respect to the deprotonation than Ag(100), as demonstrated in Supplementary Figure S2.2, where one can see that BDA on Ag(100) reaches the same degrees of the deprotonation at lower annealing temperatures than on the Ag(111). No deprotonation of BDA on Au substrate has been observed.[33–35] Hence, substrate material has a decisive role in the deprotonation of deposited carboxylic acids. Moreover, surface plane orientation also affects the reactivity and structure of molecular phases.

On both Ag(111) and Ag(100) surfaces, the structure of the intact and fully deprotonated phases is similar. The intact phases $\alpha^{(111)}$ and $\alpha^{(100)}$ both display row-like structure. Although BDA is only van der Waals bonded to the substrate, the substrate already significantly influences the resulting BDA structure. The basic motif, BDA rows, is observed on both surfaces. On the Ag(100), the interchain spacing fits the periodicity of 2 substrate atoms in the principal surface direction, making this periodicity dominant. On the other hand, the BDA molecules within the rows were not periodically positioned as it is impossible to satisfy the condition of optimal intermolecular bonding and optimal molecule substrate bonding simultaneously for all molecules within the rows. On Ag(111), the rows are rotated by 12° with respect to the principal substrate directions. This results in a slight variation in interrow spacing, but the molecules remain periodically positioned within the rows. On both surfaces, the deprotonated molecules can be incorporated[16,31,32,38] within the initial chain structure, causing slight structural changes.

For the fully deprotonated phases, the binding motif is essentially the same on both substrates. The mixed phases that feature intact, semi-, and fully deprotonated molecules significantly vary in their structure and chemical composition. A substantial difference is an absence of $\gamma^{(100)}$-like phases on Ag(111) that have been associated with $k$-uniform tilings.[32] This highlights the importance of surface in the synthesis of complex molecular structures.

**Phase transformations**

### The $\alpha^{(111)} \rightarrow \alpha_s^{(111)}$ phase transformation

A gradual increase in temperature induces the deprotonation of BDA molecules, which, in turn, causes their rearrangement and the formation of the new molecular phase. The snapshots depicted in Figure 4a were taken during the transformation from the as-deposited $\alpha^{(111)}$ to the $\alpha_s^{(111)}$ phase observed at 373 K in the LEEM bright-field mode. In contrast to the Ag(100), the $\alpha^{(111)} \rightarrow \alpha_s^{(111)}$ transformation proceeds within the existing islands, i.e., it does not involve a dissolution of the preceding phase. It develops by incorporating carboxylate groups into the

$\alpha^{(111)}$ as defects. When the concentration of carboxylates is sufficiently high, the carboxylate-substrate interaction becomes dominant, and the phase transformation to the $\alpha_s^{(111)}$ occurs. The deprotonation also continues even further within the $\alpha_s^{(111)}$ but no additional structural changes are observed. During the transformation, the smaller islands decrease in size and eventually disappear (Figure 4a), while the bigger ones gradually grow; the final mean island size is more than 4 times larger compared with the $\alpha^{(111)}$ (see above). This phenomenon can be explained by Ostwald ripening already observed at Ag(100).[31] The transformation inside the islands proceeds in a similar way as observed during the $\alpha^{(100)} \to \dot{\alpha}^{(100)}$ irreversible transition on Ag(100) in the full ML regime. This internal transformation takes place when structural motifs of the new phase can be incorporated into the structure of the previous one, as mentioned in the section describing the $\alpha_s^{(111)}$ phase. To summarize, in comparison to the $\alpha^{(100)} \to \dot{\alpha}^{(100)}$ on Ag(100), the $\alpha^{(111)} \to \alpha_s^{(111)}$ transformation at the Ag(111) reaches lower degrees of the deprotonation at the same annealing temperatures (Figure S2.2 in Supplementary Information).

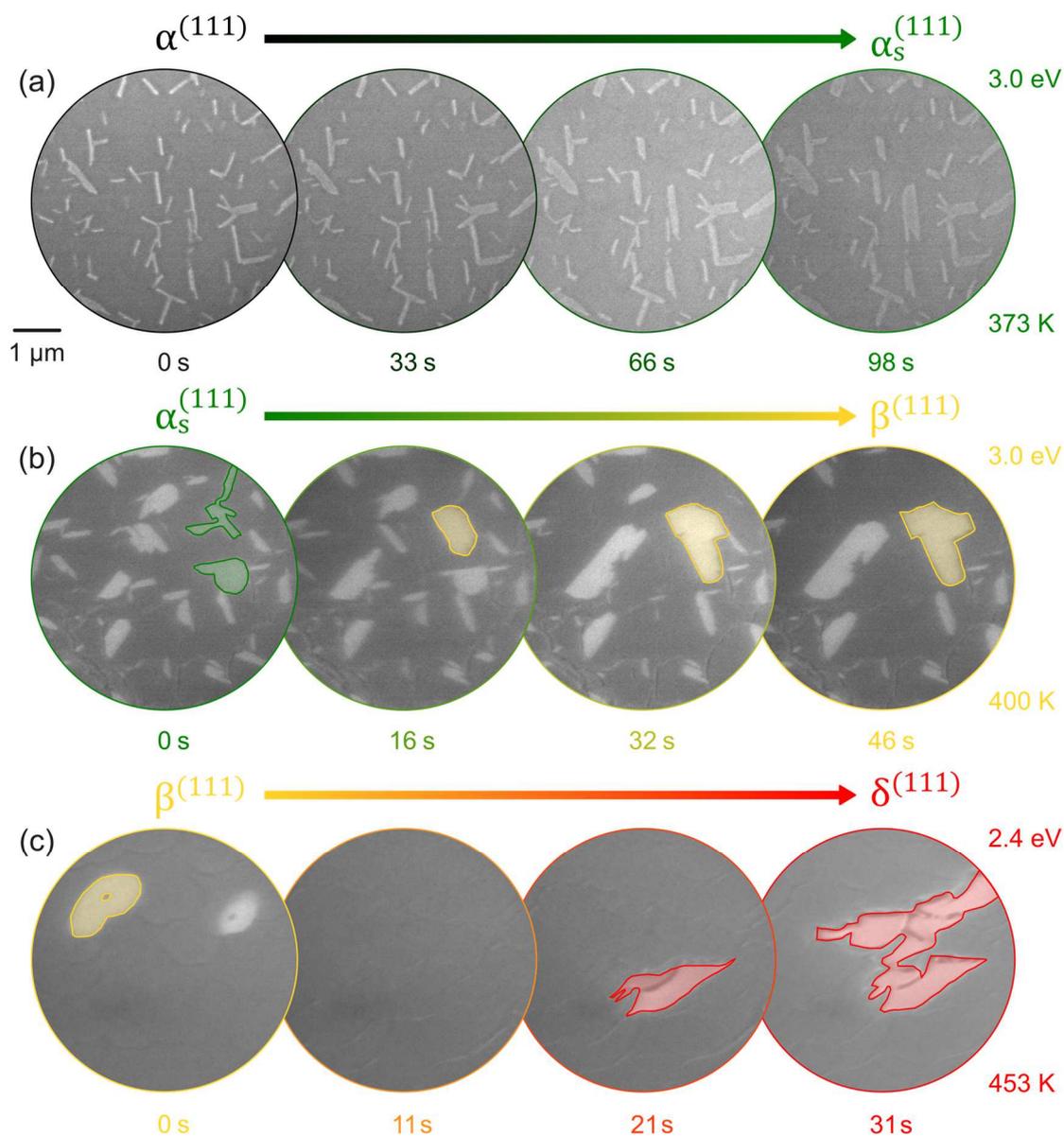

**Figure 4.** Snapshots documenting the phase transformations of BDA on the Ag(111) measured by LEEM. (a) The $\alpha^{(111)} \to \alpha_s^{(111)}$ transformation. (b) The $\alpha_s^{(111)} \to \beta^{(111)}$ transformation: The green and yellow areas highlight the $\alpha_s^{(111)}$ and $\beta^{(111)}$ phases, respectively. (c) The $\beta^{(111)} \to \delta^{(111)}$ transformation: the yellow and red areas highlight the $\beta^{(111)}$ and $\delta^{(111)}$ phases, respectively; the bright features represent the BDA molecular islands on the dark Ag(111) substrate for all the above-mentioned transformations. The full videos are provided as Supplementary Videos SV1-SV3, which are described in Supplementary Information, Section 7.

## The $\alpha_s^{(111)} \to \beta^{(111)}$ phase transformation

The $\alpha_s^{(111)} \to \beta^{(111)}$ transformation (Figure 4b) proceeds in the same manner as the $\alpha^{(100)} \to \beta^{(100)}$ does on Ag(100).[31] The nucleation of the $\beta^{(111)}$ phase happens at random places on the surface outside of $\alpha_s^{(111)}$ molecular islands; once the $\beta^{(111)}$ is nucleated, the transformation itself takes less than ~1 min at 400 K. The relatively long time (~10 min) before the nucleation[31] and fast transformation indicate the burst nucleation mechanism.[42] The transformation is accompanied by the remote dissolution of the preceding phase; the $\alpha_s^{(111)}$ islands are quickly dissolved in the vicinity of $\beta^{(111)}$ islands. Step annealing at increasing temperatures followed by synchrotron radiation photoelectron spectroscopy (Supplementary Information, Section 2) reveals that there is first a gradual increase in deprotonation (see the phase transformation of $\alpha^{(111)} \to \alpha_s^{(111)}$). When the amount of deprotonated carboxylic groups reaches ~30 %, there is a sudden increase in the deprotonation rate. The BDA molecules quickly deprotonate until ~63 % of the deprotonation is reached. We ascribe this rapid increase in the deprotonation rate to the onset of the $\beta^{(111)}$ phase formation, which further facilitates the deprotonation process. We explain this process in terms of the burst transformation (for more details see the discussion below).

## The $\beta^{(111)} \to \delta^{(111)}$ phase transformation

The $\beta^{(111)}$ phase is exceptionally stable (Figure 5) in a wide range of temperatures 403 – 453 K, where just slight deprotonation occurs. At 453 K, the $\beta^{(111)}$ quickly dissolves into the molecular gas (Figure 4c). If the heating is stopped at this point, $\delta^{(111)}$ islands nucleate and grow (Figure 4c). If the heating continues, BDA molecules decarboxylate or desorb, which eventually results in a low coverage of polyphenyls on the Ag(111) surface. Only when the

β$^{(111)}$ dissolves into the molecular gas (condensed-to-2D-gas phase transition), the deprotonation of BDA molecules beyond 63 % becomes possible. Without the evaporation step, the formation of the δ$^{(111)}$ was only rarely observed at coverages close to 1 ML.

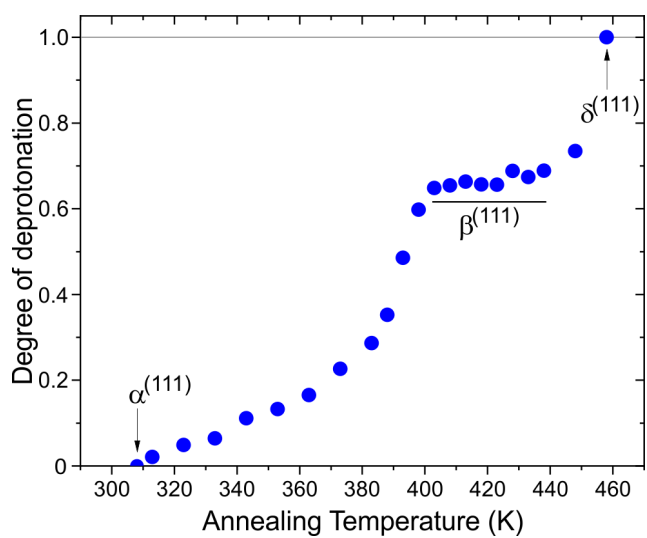

**Figure 5.** Degree of deprotonation of carboxyl groups determined from the fitting of O 1s peak measured by synchrotron radiation. The sample was gradually annealed up to 473 K. The spectra were measured after each annealing step with turned-off heating (Supplementary Information, Section 2).

**Discussion of Phase Transformations**

The phase transformations of BDA presented here and in our previous works[31,32,38] enable us to draw general conclusions. We have used the term transformation to highlight that the presented phase transitions are irreversible. The irreversibility is a direct consequence of associative hydrogen desorption from the surface under UHV conditions.[15,43] The extremely low hydrogen partial pressure in the surrounding environment causes the reverse process, i.e., dissociative hydrogen adsorption, to occur very rarely. Thermodynamically, this is expressed by a large entropy gain of 0.5 eV per H atom associated with the desorption process.[43]

The as-deposited molecules comprise two intact carboxyl groups, one or both of which can be deprotonated, rendering them chemically different. As a result, we obtain up to 3 distinct molecules that form complex assemblies. Also, several distinct types of transitions occur as a response to the thermally induced deprotonation. First order transitions were observed for $\alpha^{(100)} \rightarrow \beta^{(100)}$ at Ag(100)[31] and $\alpha_s^{(111)} \rightarrow \beta^{(111)}$ at Ag(111). In both cases, the structure of the new phase is incompatible with the previous one, and the islands of the new phase ($\beta^{(100)}$ or $\beta^{(111)}$) nucleate at free areas outside the islands of the previous phase ($\alpha^{(100)}$ or $\alpha_s^{(111)}$). The situation is slightly changed for transformations between the $\gamma^{(100)}$ phases (these were denoted 3U and 2U in the original paper) at Ag(100)[32]. The structure of the $\gamma^{(100)}$ phases share a common structural feature: all molecules in the direction of one unit cell vector have the same position and orientation. Hence the transformation between these phases can proceed locally in the voids within the islands or at their periphery. The transformation thus proceeds mainly within the voids propagating through the molecular islands while these are preserved.[32] In contrast, the $\alpha^{(111)} \rightarrow \alpha_s^{(111)}$ and $\alpha^{(100)} \rightarrow \dot{\alpha}^{(100)}$ transformations are gradual and, hence, can be termed as second-order irreversible transitions. As the deprotonated groups can be incorporated within the structural motif of the initial phases ($\alpha^{(100)}$ or $\alpha^{(111)}$), the phases seamlessly and continuously transform into new ones without nucleation. The structure of the new phases differs only slightly, i.e., the position of the individual molecules is adjusted to optimize bonding with a given surrounding.

The 2D molecular gas present between the condensed phase islands plays an indispensable role in the transformations as it mediates the mass transport between the islands.[31] Notably, the deprotonation is much faster in 2D gas than in the condensed phase. The driving force of the deprotonation is the free energy change between the initial state – the protonated carboxyl group in the vicinity of the metal substrate – and the final state in which carboxylate oxygen is bound to the substrate and the hydrogen molecule in the gas phase; the entropy difference between these two states largely contributes to the free energy balance.[43] However, in many cases, the binding to the substrate is energetically favorable even without hydrogen desorption,[44] which makes deprotonation kinetically accessible. Whereas the deprotonation can also proceed within the condensed phase, the intermolecular bonds stabilize the initial state, which increases the activation energy for deprotonation. Therefore, the deprotonation rate is much lower in the condensed phase compared with the individual molecules in the 2D gas.

The results presented here show that the $\beta^{(111)}$ phase on Ag(111) displays a fixed degree of deprotonation in a broad temperature range. There is rapid deprotonation preceding the $\beta^{(111)}$ phase formation and rapid deprotonation during the phase transition into 2D molecular gas. These results point to hindered deprotonation within the $\beta^{(111)}$ phase, i.e., stabilization of the deprotonation state within the condensed phase. An even broader temperature range is observed for the $\beta^{(100)}$ phase on the Ag(100) surface, as shown in Figure S2.2 in Supplementary Information, Section 2.

The discussion above allows us to develop the term burst transformation in terms of a combination of burst nucleation, stabilization of deprotonation state within the condensed phase, and rapid deprotonation on a bare substrate. In our preceding work,[31] we have identified that the $\alpha^{(100)} \rightarrow \beta^{(100)}$ phase transformation can be described in the framework of La Meer burst nucleation. The mechanism involves a slow increase in supersaturation, which reaches the level allowing nucleation at a certain moment. At this moment, new islands nucleate quasi-simultaneously. The presence of islands rapidly decreases the supersaturation: the molecules from the 2D gas are attached to the newly nucleated islands. Without significant supersaturation, no additional nucleation occurs. The $\beta^{(111)} \rightarrow \delta^{(111)}$ phase transformation gives additional hints on the role of the stabilization of the deprotonation state in the previous phase.

The general picture of burst transformation is thus as follows. In the beginning, there are islands of the initial phase in equilibrium with 2D molecular gas. During the annealing, with the steady increase in temperature, more and more molecules are released to the 2D gas, where they can deprotonate. This gradually builds up the 2D molecular gas, and continuing deprotonation makes its composition different from the parent phase. The equilibrium condition requires more molecules to be detached and eventually deprotonated. At some point, there is sufficient supersaturation allowing the formation of the new phase – its formation follows the burst nucleation mechanism described above. This is followed by a rapid decrease of 2D gas concentration, below the equilibrium with the previous phase. Consequently, many molecules are released from the preceding phase; these molecules are rapidly deprotonated and incorporated into new phase islands. This process, i.e., remote dissolution, continues until the islands of the initial phase are dissolved. Phase stabilization plays an important role in delaying the phase transformation as it prevents deprotonation and keeps the supersaturation low, and, consequently, nucleation cannot occur.

CONCLUSIONS

We have described the self-assembled phases of BDA on a Ag(111) surface by a set of experimental techniques: LEEM, LEED, STM, and XPS. The comparison with our earlier works on BDA on Ag(100) allows us to describe the influence of the substrate on the structure, chemical composition, and deprotonation-induced transformation of the self-assembled phases. The basic structural motifs are similar in the case of the initial phase comprising only intact BDA molecules and the final phase containing only fully deprotonated molecules. Here only minor differences are observed, i.e., the distinct orientation of the molecular rows in the intact phases ($\alpha^{(100)}$ and $\alpha^{(111)}$) or the different density of the final phases ($\delta^{(100)}$ and $\delta^{(111)}$). The initial $\alpha^{(100)}$ and $\alpha^{(111)}$ phases can incorporate the deprotonated carboxyl groups, first as structural defects and, at a higher concentration, as structural units of the new $\alpha_s^{(111)}$ and $\dot{\alpha}^{(100)}$ phases. The other phases – $\beta^{(100)}$, $\gamma^{(100)}$, and $\beta^{(111)}$ – are different between the substrates.

The transformations from β phases reveal an important new feature: the stabilization of the chemical state of the molecules within the condensed phase. The non-α phases possess a given structure in which the intact, partially, and fully deprotonated molecules coexist in a given ratio. The deprotonation state in these condensed phases is quite stable, and consequently, these phases are stable in a broad range of temperatures. The phase stabilization and the previously identified burst nucleation combine to the burst transformation scheme discussed here.

In conclusion, our study highlights the necessity of detailed studies of the structure and behavior of molecular layers on surfaces with distinct orientations for developing viable functionalization strategies of substrates displaying multiple surface planes (e.g., nanocrystals, polycrystalline foils), as was shown in this work for BDA on the Ag surfaces. For these systems, we have provided a generalized description of the kinetics of the molecular phase transitions near thermodynamic equilibrium.

ASCIATED CONTENT

**Supporting Information**.

The following files are available free of charge.

Description and discussion on (1) XPS analysis of the BDA phases on the Ag(111); (2) Synchrotron radiation photoelectron spectroscopy of the BDA phases on the Ag(111); (3) LEEM/LEED analysis of all the phases; and (4) Phase transitions measured by LEEM. (PDF)

LEEM videos: alpha-s_373-374K_20f-s.avi; beta_400K_10f-s; delta_453K_22font_10f-s (AVI)

ACKNOWLEDGMENT

This research has been supported by GAČR (grant number: 22-05114S). CzechNanoLab project (grant number: LM2018110) funded by MEYS CR is gratefully acknowledged for the financial support of the measurements at CEITEC Nano Research Infrastructure. The authors acknowledge the CERIC-ERIC Consortium for access to experimental facilities and financial support.

AUTHOR INFORMATION

**Corresponding Author**

* E-mail: [cechal@fme.vutbr.cz](mailto:cechal@fme.vutbr.cz) (J. Č.)

**Author Contributions**

J.Č. provided the idea and conceptualized the research. A.M., L.K., P.P., and M.B. performed the laboratory experiments. P.P, A.M., T.S., M.B., and J.Č. measured synchrotron radiation data. J.Č. supervised the experiments and developed the methodology. A.M., P.P., M.B., and J.Č. analyzed the data. A.M., M.B. and J.Č. wrote the manuscript.

ADDITIONAL INFORMATION

Authors declare no competing financial interests.

TOC graphics

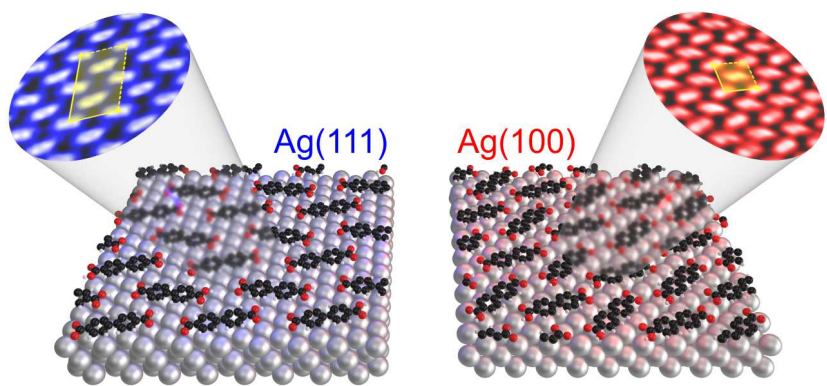

# The Role of Phase Stabilization and Surface Orientation in 4,4'-Biphenyl-Dicarboxylic Acid Self-Assembly and Transformation on Silver Substrates

*Anton O. Makoveev,[1] Pavel Procházka,[1] Matthias Blatnik,[1] Lukáš Kormoš,[1] Tomáš Skála,[2] and Jan Čechal[1,3]\**

[1] CEITEC - Central European Institute of Technology, Brno University of Technology, Purkyňova 123, 612 00 Brno, Czech Republic.

[2] Department of Surface and Plasma Science, Faculty of Mathematics and Physics, Charles University, V Holešovičkách 2, 18000 Prague 8, Czech Republic

[3] Institute of Physical Engineering, Brno University of Technology, Technická 2896/2, 616 69 Brno, Czech Republic.

CONTENTS:

# 1. XPS analysis of the BDA phases on Ag(111)

The O 1s spectra were fitted by 2 – 3 Voight components (Lorentzian width 0.1 eV) and a combined Shirley–parabolic background, as shown in Figure 2 of the main text. The parabolic background component was introduced to model a decreasing background intensity at the high binding energy side of the Ag 3d peak. The peak-fitting parameters are summarized in Table S1, and all of them are consistent with those for BDA on the Ag(001) substrate and spectra measured using synchrotron radiation (Section 2 of this Supplementary Information).

## Analysis of the $\alpha^{(111)}$ phase

The O 1s spectrum can be deconvoluted into two peaks: one at 533.7 eV (magenta in Figure 2 in the main text) is ascribed to hydroxyl oxygens, and the second at 532.35 eV (blue) is to carbonyl oxygens [1,2]. Their signal intensities have a ratio of 1:1, which corresponds to an expected equal number of hydroxyls and carbonyls in the fully protonated molecules. The peak positions on Ag(111) are shifted by 0.1 eV to lower binding energies (BE) compared to the one on Ag(100).

## Analysis of the $\alpha_s^{(111)}$ phase

The XPS spectra shown in Figure 2 in the main text depict two possible states of the $\alpha_s^{(111)}$ with different degrees of BDA deprotonation. As in the case of the $\alpha^{(111)}$ phase, the magenta, and blue peaks (Figure 2 in the main text) belong to the hydroxyl and carbonyl oxygens, respectively. The orange peak at 530.9 eV indicates oxygen in the COO⁻ group. Reasonable fitting of the spectra can be obtained only when the carboxyl-carboxylate motif is considered, which requires two additional peaks associated with hydroxyl and carbonyl in carboxyl bound to carboxylate (violet and dark blue peak components in in Figure 2 in the main text). This is in accordance with the analogous phases $\acute{\alpha}^{(100)}$ and $\beta^{(100)}$ on Ag(100). The laboratory XPS measurements are confirmed by the high-resolution SRPES investigation (Section 2). The BEs of the peaks obtained by the SRPES analysis are 533.75, 532.35, and 530.75 eV for the hydroxyl, carbonyl oxygens in carboxyl groups, and carboxylate oxygens, respectively. This corresponds very well with the measurements from our laboratory XPS instrument.

Table S1.1: Peak-fitting parameters: binding energy position (BE) and FWHM of the Gaussian part of the O 1s peak components. The BE shifts (from the peak, marked +0.0) are given in brackets. The fraction of deprotonated carboxyl groups is given for all the phases

|  | Component O1 (hydroxyl oxygen) | | Component O2 (carbonyl oxygen) | | Component O3 (carboxylate oxygen) | | Deprotonation |
|---|---|---|---|---|---|---|---|
|  | BE (eV) | FWHM (eV) | BE (eV) | FWHM (eV) | BE (eV) | FWHM (eV) |  |
| $\alpha^{(111)}$ | 533.7 (+1.35) | 1.7 | 532.35 (+0.0) | 1.6 | – | – | 0.0 |
| $\alpha_s^{(111)}$ | Combined α and β fit | | | | | | 0.1 – 0.3 |
| $\beta^{(111)}$ | 532.65 (+0.95) | 1.6 | 531.7 (+0.0) | 1.6 | 530.9 (−0.8) | 1.5 | 0.63 |
| $\delta^{(111)}$ |  |  |  |  | 530.45 | 1.3 | 1.0 |

## 2. Synchrotron radiation photoelectron spectroscopy of the BDA phases on Ag(111)

The O 1s spectra were fitted by 2 – 3 Voight components and a Shirley background, as shown in Figure S1; the peak fitting parameters are summarized in Table 2. All fits are fully consistent with our laboratory measurements using a standard, non-monochromatic X-ray source and our previous results on the Ag(001) substrate. For the C 1s spectra, the synchrotron radiation provides a much higher resolution than it is achievable in our laboratory. This enabled to distinguish the split of carboxyl related C 1s component (C2) in the $\beta^{(111)}$ phase and better describe shake-up satellites. The C 1s peak parameters are summarized in Table 3.

Consistently with laboratory-based measurement, a shoulder at the higher BE side of the main component was necessary to include to obtain a good fit. The dominant feature of the C 1s spectrum is the existence of an extensive shake-up satellite structure; we have modeled it by several broad peaks shown in dark yellow in Figure S1. These peaks complicate obtaining quantitative results from the fitting procedure if they overlap with other peaks. This is the case for the spectrum taken after the annealing at 420 K (the $\beta^{(111)}$ phase). In this phase, the C2 peak has two subcomponents, one at higher BE (C2H) and the second at lower BE (C2L). In the $\beta^{(111)}$ phase, one of the satellites has the same BE and comparable intensity as C2H peak. The presence of the satellite can be inferred from (i) the presence of the satellite at this position in both $\alpha^{(111)}$ and $\delta^{(111)}$ phases (cf., Figure S1) and (ii) the intensity of carboxyl carbon peaks. Whereas the first point strongly supports its presence also in the case of $\beta^{(111)}$, the second point gives us a quantitative tool to split intensity between the C2H component and the satellite peak.

Without a satellite, the intensity ratio of carboxyl and phenyl associated peaks C2/C1 would be 0.22, significantly higher than the value expected from the number of carbon atoms in the BDA molecule (two carboxyl and 12 phenyl carbon atoms, i.e., C2/C1 = 2/12 = 0.165). Hence, we have decreased the C2H component intensity accordingly. Doing so, we obtain the ratio of C2H and C2L intensities 0.6:1.0; i.e., the C2H has 37% of intensity (0.6/1.6) and C2L 63 % of total C2 component intensity. This division matches with a fraction of deprotonated carboxyl groups in the $\beta^{(111)}$ phase of 5/8 = 63 %. Hence, the C2H is probably associated with protonated carboxyl groups and C2L with deprotonated carboxyl ones. This association is also consistent with BE position of C2H and C2L components. The shift of C2L in respect to the main component is 3.15 eV is almost the same as in the fully deprotonated delta phase (3.1 eV), which features only fully deprotonated carboxyl groups. In the case of C2H, the shift is 4.0 eV which is lower than in the $\alpha^{(111)}$ phase (4.4 eV). However, we should consider that C2H peak is associated with the carboxyl–carboxylate bonding instead of carboxyl–carboxyl bonding. In the first case, the carboxyl group is bonded to partially negatively charged carboxylate, and charge redistribution thus induces the decrease in the C2H BE.

**Table S2.1:** Peak-fitting parameters: BE and FWHM of the Gaussian part of the O 1s peak components. Voigt function was used for fitting; the width of the Lorentzian component was 0.1 eV. The BE shifts (from the peak, marked +0.0) are given in brackets. The fraction of deprotonated carboxyl groups is given for all the phases.

|  | Component O1 (hydroxyl oxygen) | | Component O2 (carbonyl oxygen) | | Component O3 (carboxylate oxygen) | | Deprotonation |
| --- | --- | --- | --- | --- | --- | --- | --- |
|  | BE (eV) | FWHM (eV) | BE (eV) | FWHM (eV) | BE (eV) | FWHM (eV) |  |
| **Multilayer** | 534.20 (+1.4) | 1.4 | 532.80 (+0.0) | 1.2 | – | – | 0.0 |
| **$\alpha^{(111)}$** | 533.75 (+1.4) | 1.4 | 532.35 (+0.0) | 1.4 | – | – | 0.0 |
| **$\beta^{(111)}$** | 532.45 (+0.9) | 1.4 | 531.55 (+0.0) | 1.4 | 530.75 (−0.8) | 1.3 | 0.63 |
| **$\delta^{(111)}$** |  |  |  |  | 530.35 | 1.1 | 1.0 |

Positions of Ag $3d_{5/2}$ peak were in the interval of (368.24±0.02) eV for all the measurements.

**Table S2.2:** Peak-fitting parameters: BE and FWHM of the Gaussian part of the C 1s peak components. Voigt function was used for fitting; the width of the Lorentzian component was 0.2 eV. The BE shifts (from the peak, marked +0.0) are given in brackets. The ratio of C2 and C1 intensities matches the theoretical one (2/12 = 0.165).

|  | Component C1 (phenyl rings) | | Component C2 (carboxyl carbon) | | C2/C1 |
|---|---|---|---|---|---|
|  | BE (eV) | FWHM (eV) | BE (eV) | FWHM (eV) |  |
| $\alpha^{(111)}$ | 285.11 (+0.0) | 0.8 | 289.49 (+4.42) | 0.7 | 0.15 |
| $\beta^{(111)}$ | 284.48 (+0.0) | 0.8 | 288.48 and 287.64 (+4.00 and 3.16) | 0.9 | 0.16 |
| $\delta^{(111)}$ | 284.26 (+0.0) | 0.7 | 287.35 (+3.09) | 1.0 | 0.16 |

A shoulder for the C1 component was necessary to obtain a good fit; it was shifted by 0.7 eV to higher BE, its intensity was 0.165 of the main component, and had the same Gaussian width.

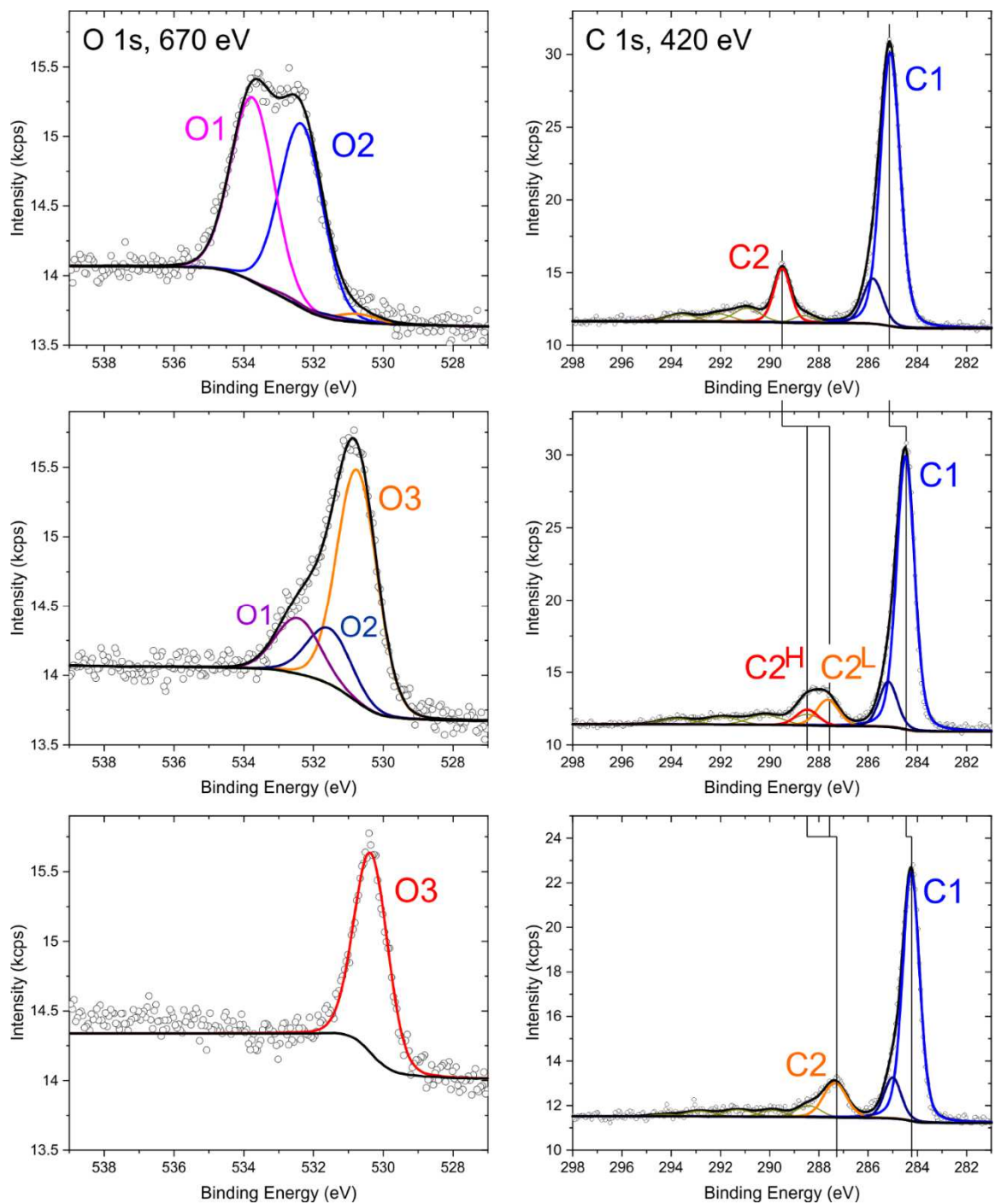

**Figure S2.1**: O 1s and C 1s spectra measured by synchrotron radiation photoelectron spectroscopy for as-deposited sample (top row), at the temperature of 400 K (middle), and 453 K (bottom).

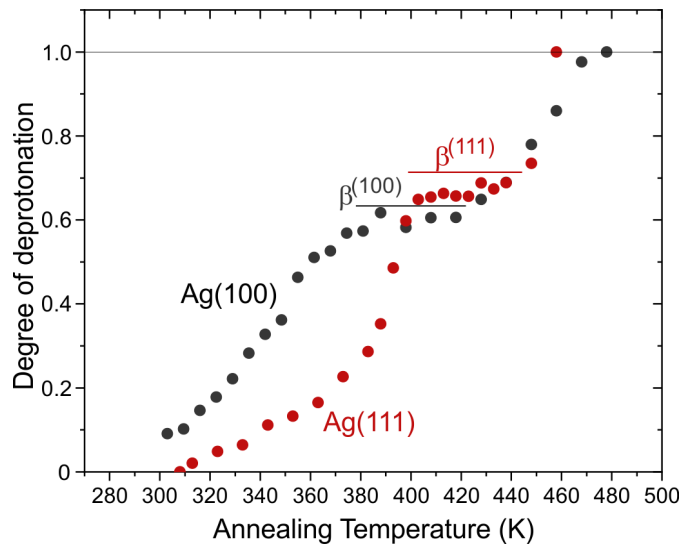

**Figure S2.2:** The degree of the deprotonation in submonolayers (0.9 ML) of BDA on the Ag(100) and Ag(111) as a function of annealing temperature. The black and red lines indicate the temperatures at which the degree of the deprotonation only slightly changes with temperatures; we associate this region with the $\beta^{(100)}$ phase. The grey horizontal line highlights the complete deprotonation, which corresponds to the $\delta^{(111)}$ and $\delta^{(100)}$ phases.

## 3. LEEM/LEED analysis of the α$^{(111)}$ phase

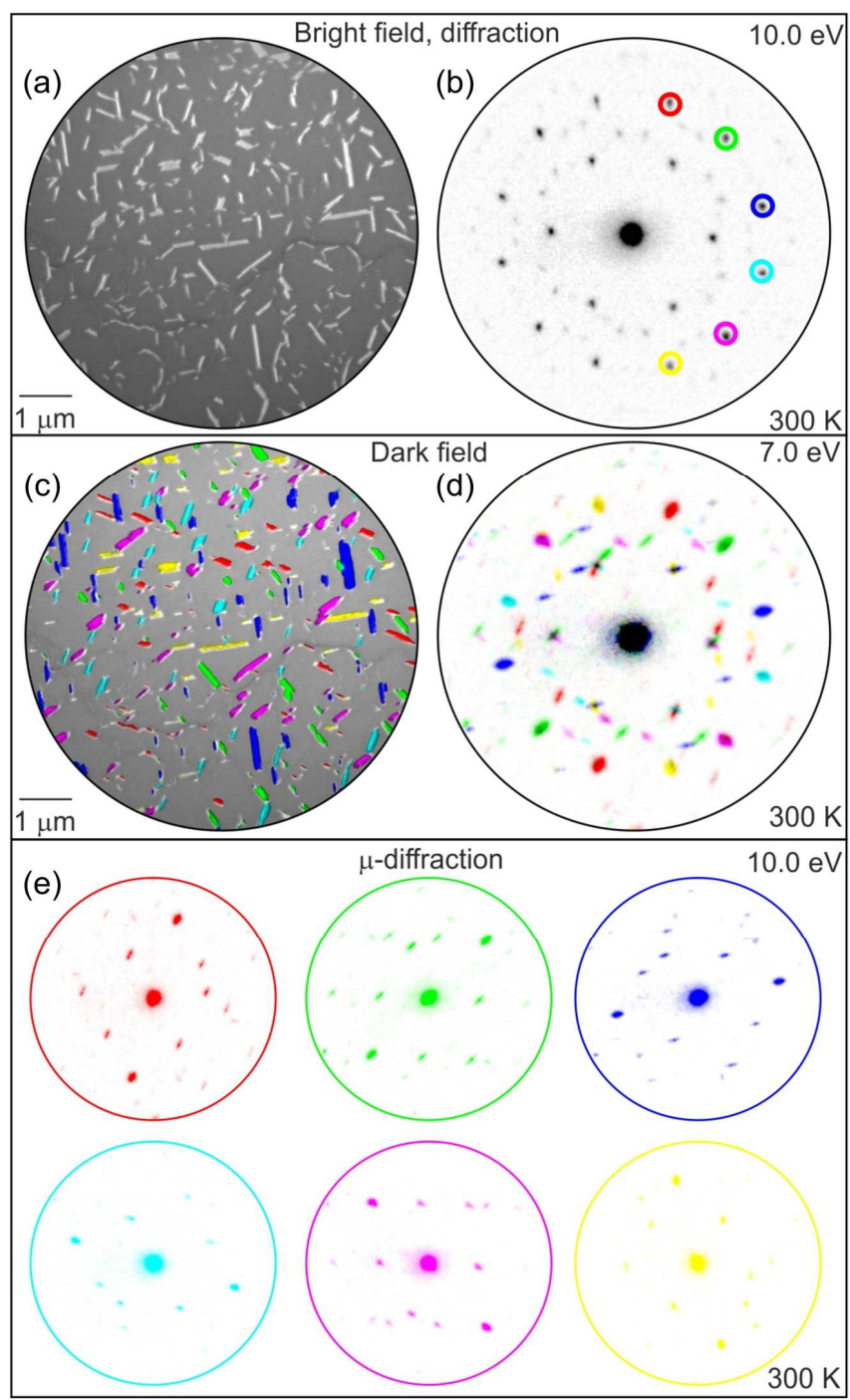

**Figure S3.1:** A LEEM/LEED analysis of the α$^{(111)}$ phase. (a) A bright-field LEEM image. The bright elongated features are BDA molecular islands on the dark Ag(111) substrate. (b) Large-scale diffraction of the α$^{(111)}$ phase. The color circles indicate the diffraction spots which belong to individual rotational domains of the α$^{(111)}$ on the Ag(111) and have been used to measure dark-field LEEM. The large-area (measured from ~118 µm²) diffraction discloses 6 symmetry-equivalent rotational domains, which are expected for an oblique unit cell on the Ag(111) substrate with p6mm symmetry. (c) A color representation of the BDA islands based on the dark-field LEEM measured form the diffraction spots designated by the color circles in (b). The combination of dark-field LEEM images and microdiffraction patterns reveals that the molecular islands are elongated in the same direction as the molecular rows comprising them. (d) A combination of the microdiffraction patterns from (e). (e) Microdiffraction of six $\alpha^{(111)}$ rotational domains on the Ag(111). One can notice an elongation of the diffraction spots. Based on our previous works,[1,3] we relate this to a slight variation of the inter-row spacing.

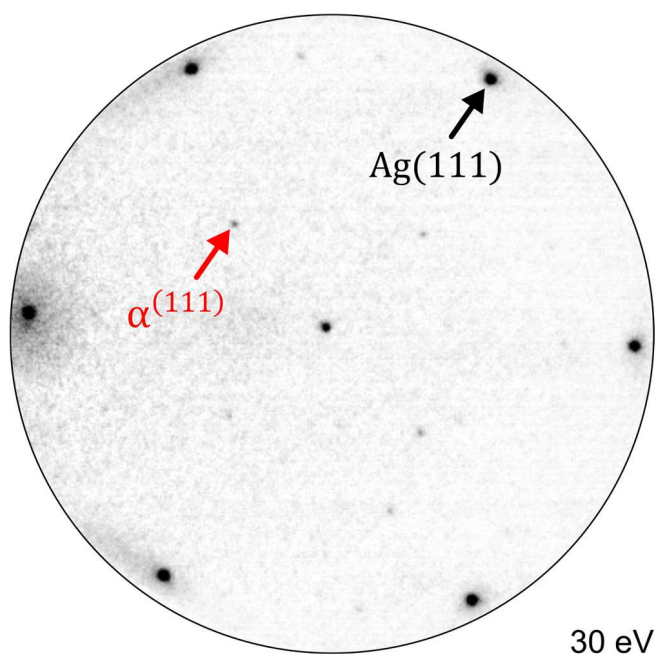

**Figure S3.2:** An $\alpha^{(111)}$ microdiffraction pattern measured at 30 eV.

## 4. LEEM/LEED and STM analysis of the $\alpha_s^{(111)}$ phase

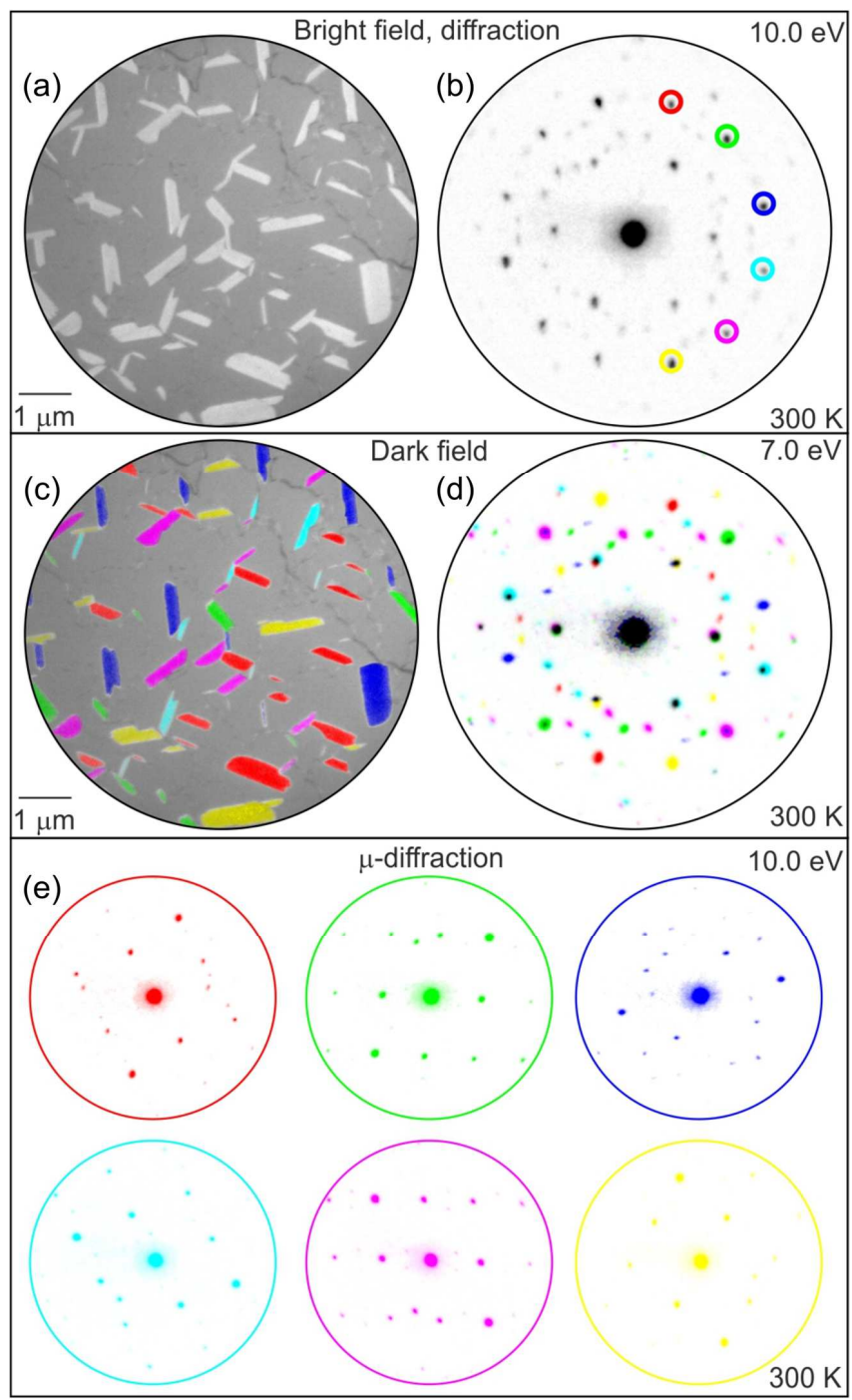

**Figure S4.1:** A LEEM/LEED analysis of the $\alpha_s^{(111)}$ phase. (a) A bright-field LEEM image. The bright elongated features are BDA molecular islands on the dark Ag(111) substrate. (b) Large-scale diffraction of the $\alpha_s^{(111)}$ phase. The color circles indicate the diffraction spots which belong to individual rotational domains of the $\alpha_s^{(111)}$ on the Ag(111) and have been used to measure dark-field LEEM. (c) A color representation of the BDA islands based on the dark-field LEEM measured form the diffraction spots designated by the color circles in (b). (d) A combination of the microdiffraction patterns from (e). (e) Microdiffraction of six $\alpha_s^{(111)}$ rotational domains on the Ag(111).

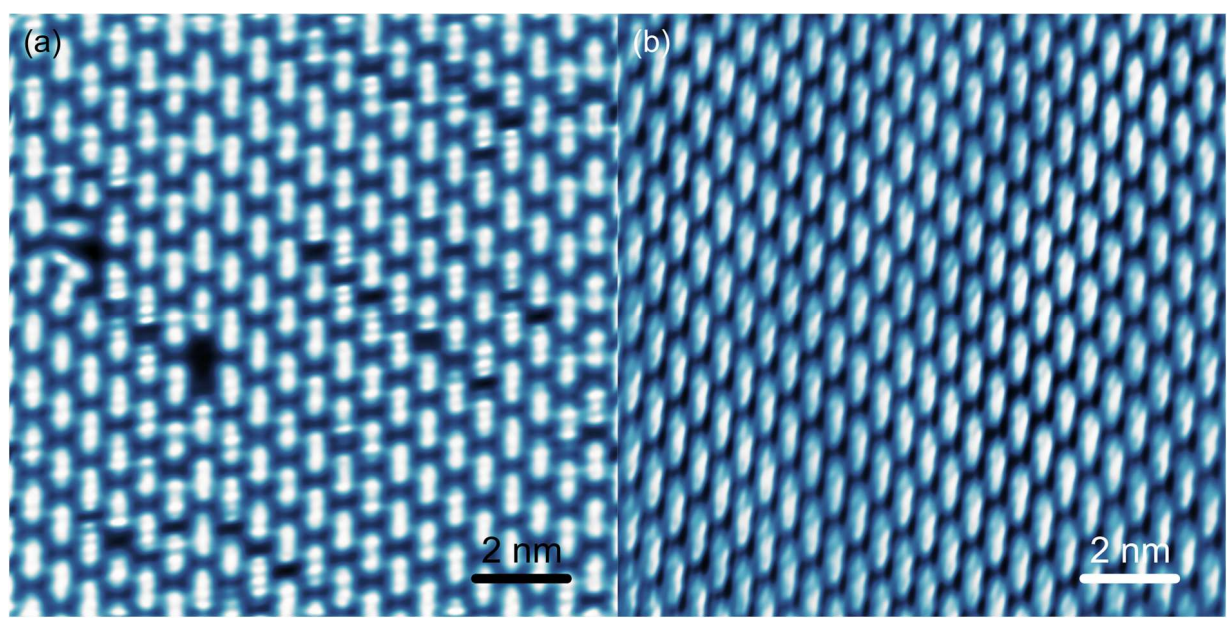

**Figure S4.2:** Larger scale STM of the $\alpha^{(111)}$ (a) and $\alpha_s^{(111)}$ (b). The corresponding measurement parameters are $U = -350$ mV and $I = 150$ pA for (a) as well as $U = -150$ mV and $I = 160$ pA for (b). Both images have been measured in constant current mode.

## 5. LEEM/LEED analysis of the β$^{(111)}$ phase

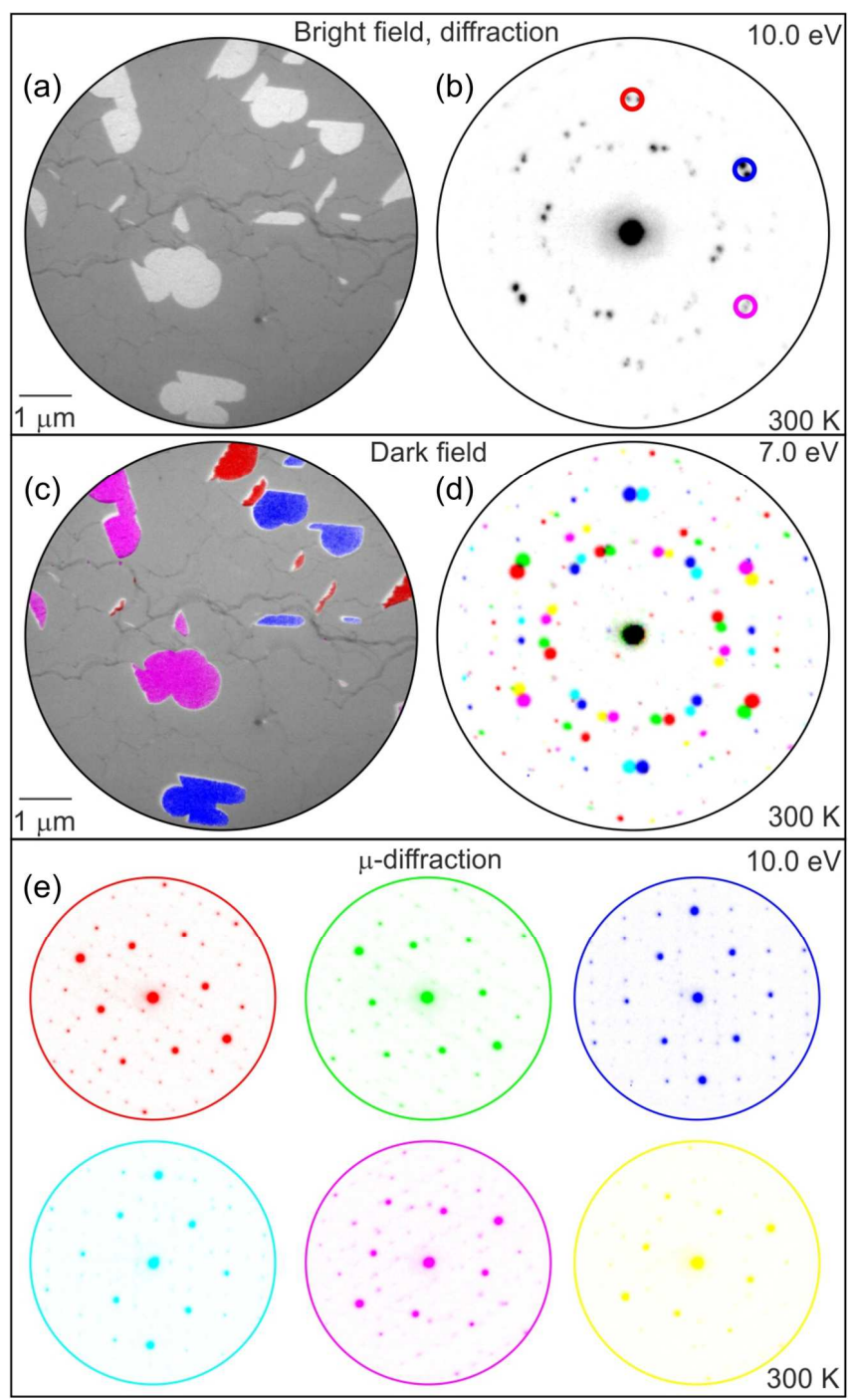

**Figure S5.1:** A LEEM/LEED analysis of the β$^{(111)}$ phase. (a) A bright-field LEEM image. The bright roundish features are BDA molecular islands on the dark Ag(111) substrate. (b) Large-scale diffraction of the β$^{(111)}$ phase. The color circles indicate the diffraction spots which belong to 3 pairs of the rotational domains of the β$^{(111)}$ on the Ag(111) and have been used to measure dark-field LEEM; in this case the diffraction spots of the different domains are so close

to each other that we could not do the separate dark-fields. (c) A color representation of the BDA islands based on the dark-field LEEM measured form the diffraction spots designated by the color circles in (b). (d) A combination of the microdiffraction patterns from (e). (e) Microdiffraction of six $\beta^{(111)}$ rotational domains on the Ag(111).

Within our experiments, we have observed two different moiré patterns of the $\beta^{(111)}$ phase (Figure S5.3). The difference is most visible when the moiré diffraction spots from the same area are compared. Figures S5.3c, d clearly show that the position of the spots in both diffraction patterns is different. An analysis of the two different diffractions reveals two slightly different BDA unit cells. In addition to the unit cell described in the main text $\begin{pmatrix} 8 & \frac{96}{11} \\ -3 & \frac{28}{11} \end{pmatrix}$, a cell with $\begin{pmatrix} \frac{97}{11} & \frac{9}{11} \\ \frac{28}{11} & \frac{61}{11} \end{pmatrix}$ dimensions coexists. Their areas are 336.56 Å² and 338.79 Å², respectively. This shows that the BDA overlayer is commensurate with the Ag(111) with a large unit cell containing ~44 molecules and several possibilities of such long-range arrangements can be found on the surface. However, the molecular unit cells (Figure S5.2) are almost identical in both cases.

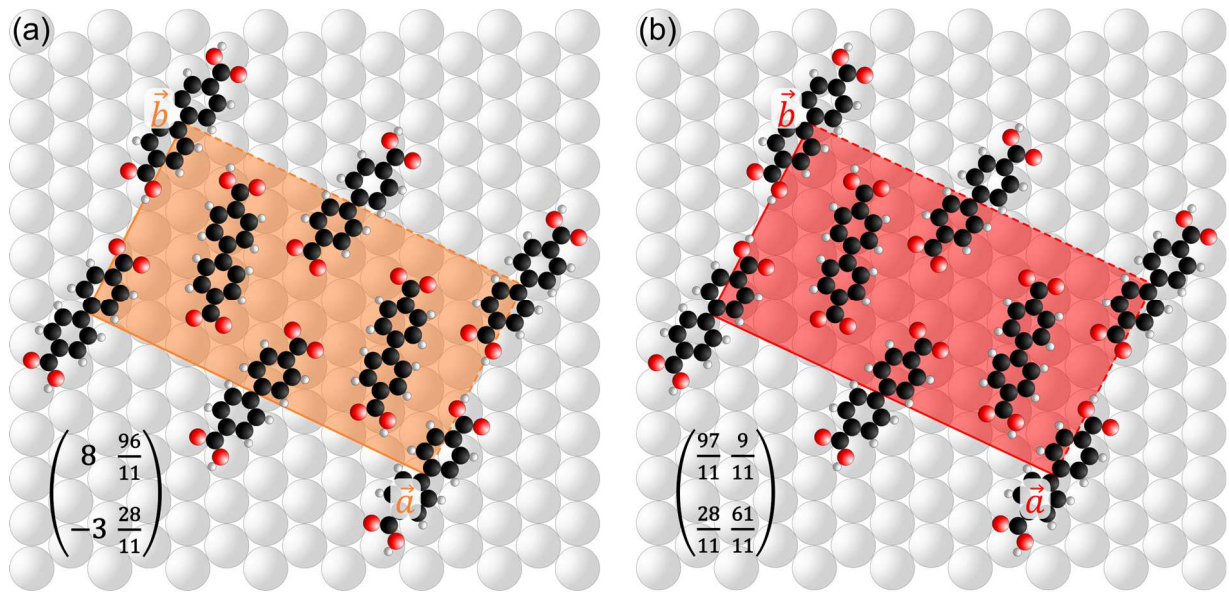

**Figure S5.2:** The two β$^{(111)}$ structures with slightly different unit cells.

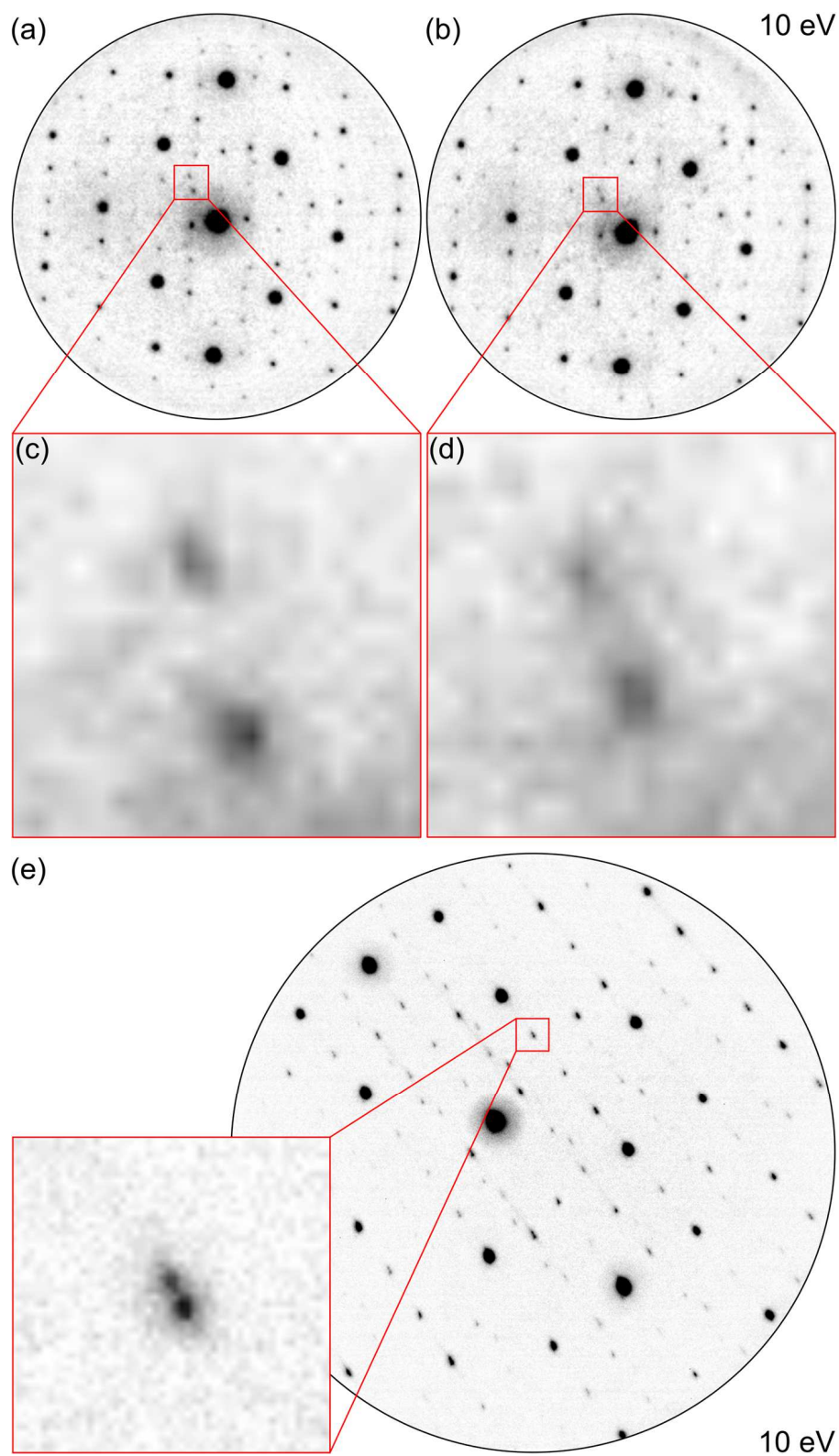

**Figure S5.3:** Two structures of the $\beta^{(111)}$ phase. (a) and (b) microdiffraction patterns of the $\beta^{(111)}$ showing different arrangements of the moiré diffraction spots. (c) and (d) zoom-ins of the same part of the reciprocal space in the (a) and (b), respectively, highlighting the arrangement difference. (e) Microdiffraction of the $\beta^{(111)}$ with the two structures (two different moiré diffraction spots arrangements) present. The inset is the zoomed in area of (e) where the existence of two distinct diffraction spots is clearly observable.

# 6. LEEM/LEED analysis of the δ$^{(111)}$ phase

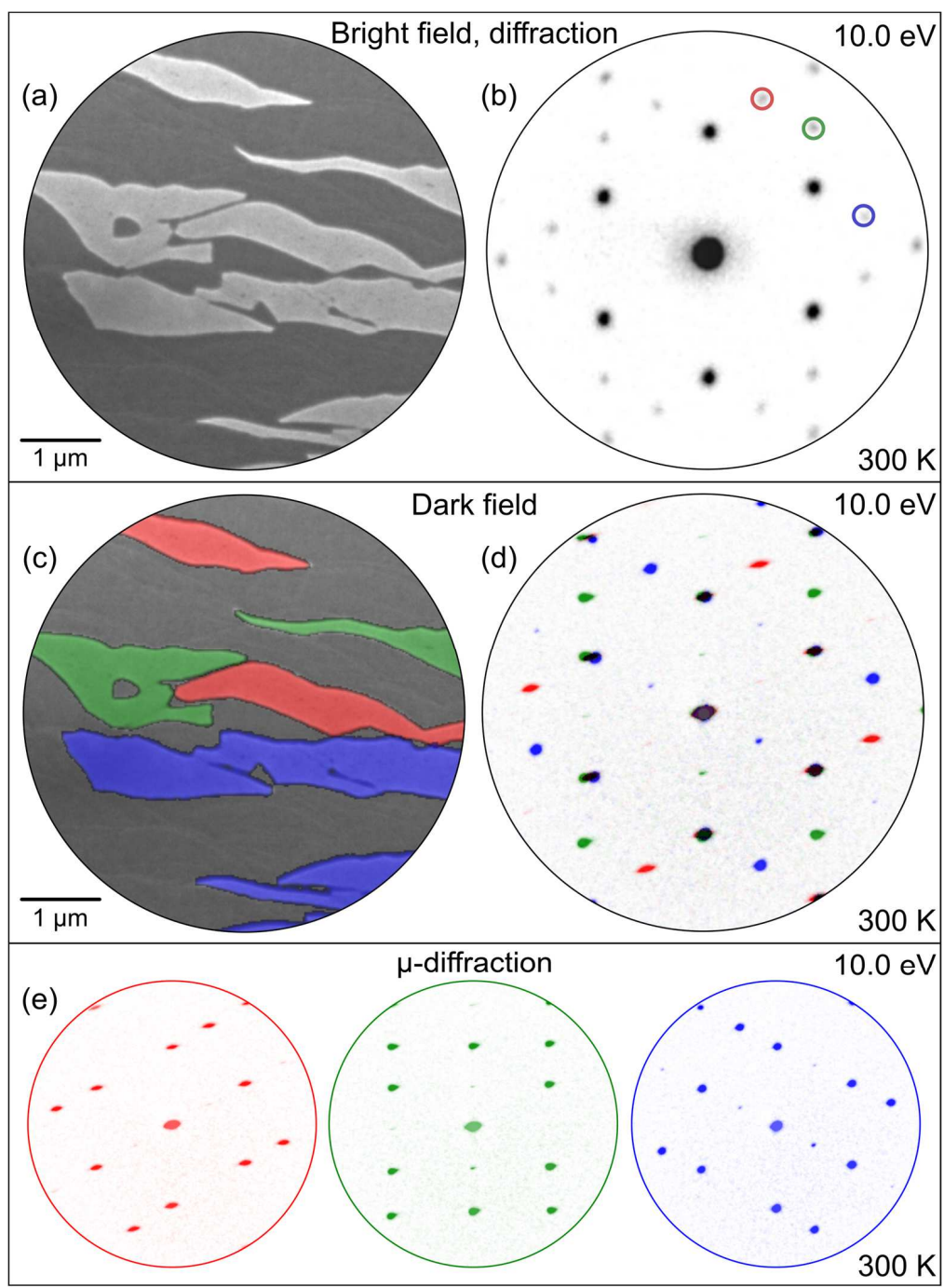

**Figure S6.1:** A LEEM/LEED analysis of the δ$^{(111)}$ phase. (a) A bright-field LEEM image. The bright features are BDA molecular islands on the dark Ag(111) substrate. (b) Large-scale diffraction of the δ$^{(111)}$ phase. The color circles indicate the diffraction spots which belong to individual rotational domains of the δ$^{(111)}$ on the Ag(111) and have been used to measure dark-field LEEM. LEED measurements and microdiffraction characterization disclose only 3 rotational

domains. The reduction in the number of symmetry-equivalent domains occurs for rectangular unit cells with one unit vector aligned with a principal surface direction; in this case, the domains that are rotated by 180° and those mirrored are symmetry equivalent. (c) A color representation of the BDA islands based on the dark-field LEEM measured form the diffraction spots designated by the color circles in (b). (d) A combination of the microdiffraction patterns from (e). (e) Microdiffraction of three $\delta^{(111)}$ rotational domains on the Ag(111). Note the systematic extinction of the diffraction spots due to presence of a glide symmetry. The spots are not completely extinct probably due to slight differences in the shape of the inequivalent molecules within the unit cell. The distinct appearance is noticeable in the STM image given in Figure 3l in the main text.

## 7. Phase transformations measured by LEEM

The $\alpha^{(111)} \to \alpha_s^{(111)}$, $\alpha_s^{(111)} \to \beta^{(111)}$ and $\beta^{(111)} \to \delta^{(111)}$ phase transformations have been measured in a bright-field LEEM mode and are demonstrated in Supplementary Videos SV1-SV3. The measurement energies of the electron beam given below have been chosen to obtain a reasonable contrast between BDA islands and the substrate and prevent undesirable effects of the beam on the molecular structures. Thus, the measurement energies for the $\alpha^{(111)} \to \alpha_s^{(111)}$, $\alpha_s^{(111)} \to \beta^{(111)}$ and $\beta^{(111)} \to \delta^{(111)}$ were 3 eV, 3 eV and 2.4 eV, respectively. The framerate of all the original measured videos is 1 frame/second. The heating rate for all the transformations was ~1 K/minute. We have prepared image sequences with the corresponding transformations out of the videos (shown in Figure 4 of the main text).